\documentclass[runningheads]{llncs}
\usepackage[T1]{fontenc}
\usepackage{ragged2e}

\usepackage{cite}
\usepackage{url}

\usepackage{textcomp}
\usepackage{gensymb}
\usepackage{graphicx}
\usepackage{epstopdf}

\usepackage{amsmath}
\usepackage{balance}

\usepackage{xcolor}
\usepackage{colortbl}

\usepackage{hyperref}
\hypersetup{
	colorlinks=true,
	linkcolor=blue,
	filecolor=magenta,      
	urlcolor=cyan,
	pdftitle={Overleaf Example},
	pdfpagemode=FullScreen,
}

\usepackage[bottom]{footmisc}
\usepackage{fancyhdr}
\usepackage{multicol}
\usepackage[
type={CC},
modifier={by},
version={4.0},
]{doclicense}

\usepackage{amssymb}
\usepackage{graphicx}
\begin{document}
\title{The generalized method of solving ECDLP using quantum annealing}
\author{\L{}ukasz Dzierzkowski\inst{1}\orcidID{0000-0002-9204-4558}}
\authorrunning{\L{}. Dzierzkowski}
\institute{Faculty of Cybernetics, Military University of Technology, Warsaw, Poland
\email{lukasz.dzierzkowski[at]wat.edu.pl}}
\maketitle              % typeset the header of the contribution
\begin{abstract}
This paper presents a generalization of a method allowing the transformation of the Elliptic Curve Discrete Logarithm Problem (ECDLP) over prime fields to the Quadratic Unconstrained Binary Optimization (QUBO) problem. The original method requires that a given elliptic curve model has complete arithmetic. The new one has no such restriction, which is a breakthrough. Since the mentioned obstacle is no longer a problem, the latest version of the algorithm may be used for any elliptic curve model. As a result, one may use quantum annealing to solve ECDLP on any given model of elliptic curves.\texttt{}

\end{abstract}
\section{Introduction}

Elliptic curves have many practical applications nowadays, such as: encryption, key exchange, digital signatures, primality testing and factorization. The first three of them require ensuring safe use, which means that any eavesdropper will not be able to recover protected data or forge a valid signature. One of the mechanisms, that allow protocols in the elliptic curve cryptography (ECC) to remain secure, is the hardness of solving the elliptic curve discrete logarithm problem (ECDLP). One can choose one of two possible ways to deal with this problem: classical or quantum.

While considering the first option, many algorithms are available. The most desired are those, which may be used in arbitrary cases, e.g.:
\begin{itemize}
	\item the method of Pohlig-Hellman \cite{pohlig1978improved},
	\item Baby step, giant step method \cite{shanks1973five, bernstein2013two},
	\item Pollard's $\rho$ method \cite{pollard1978monte},
	\item Pollard's $\lambda$ method \cite{pollard1978monte}.
\end{itemize}
They all have one thing in common -- fully exponential computational complexity in a general case \cite{washington2008elliptic}. That means, ECDLP used in modern cryptosystems cannot be solved in a reasonable time with a classical algorithm. Moreover, it is hard to notice any prospect of changing this situation, at least in the near future. If so, other ways should be considered.

The second option is quantum computing. The most known device in the described category is a general purpose quantum computer (GPQC). There is only one known algorithm, which may be used to solve ECDLP on the mentioned device -- Shor's algorithm. However, none of the ECDLP examples have been solved with this method so far. The reason for this state of affairs was mentioned by Martin Roetteler et al. in \cite{roetteler2017quantumresourceestimatescomputing}. The authors of that paper showed, what is the number of Toffoli gates required to solve ECDLP with Shor's algorithm \cite{shor1994algorithms}. For an elliptic curve on some $n$-bit prime field, it could be even $448n^3\log_2(n)+4090n^3$, which exceeds the achievable resources.

Fortunately, quantum computing does not end with GPQC. In the field of cryptography, quantum annealing (QA) is increasingly used. So far, many applications of QA in cryptanalysis have been described, i.e.:
\begin{itemize}
	\item solving discrete logarithm problem (DLP) \cite{wronski2022practical},
	\item conducting an algebraic attacks on block ciphers \cite{burek2022algebraic, burek2022quantum},
	\item conducting an algebraic attack on stream ciphers \cite{wronski2023security},
	\item integer factorization \cite{jiang2018quantum, peng2019factoring, zolnierczyk2023searching, ding2024effective},
	\item \textbf{solving ECDLP} \cite{wronski2021index, wronski2024transformation, wronski2024base}.
\end{itemize}
The last item on the list concerns the problem that is the main topic of this paper. As may be seen, solving ECDLP with QA has already been described. Nevertheless, none of the mentioned articles presents a fully-quantum method of solving the ECDLP for an arbitrary case (\cite{wronski2021index} uses both classical and quantum computations and \cite{wronski2024transformation, wronski2024base} require to conduct the operations on an elliptic curve with complete arithmetic). 

The breakthrough will be presented in this article. The method described below does not require any computations on a classical computer (just converting the ECDLP to some specific form) and may be used for any elliptic curve model, even without complete arithmetic.

\section{Theory}

The quantum annealer D-Wave, which allows building applications to access and use QA in the cloud, accepts a few different classes of problems, for example:
\begin{enumerate}
	\item Binary Quadratic Models (BQC),
	\item Constrained Quadratic Models (CQM),
	\item Discrete Quadratic Models (DQM).
\end{enumerate}
Solving ECDLP with QA requires transforming the problem to the BQM. There are two possible representations in that class: the Ising model and the quadratic unconstrained binary optimization model (QUBO). Switching between models is almost effortless. Due to practical purposes and because of the differences in the way of representing variables in both models, the QUBO model is more suitable for solving ECDLP and therefore it was used.

In the below subsections, brief descriptions of ECDLP and QUBO are provided. Then the original way of solving the described problem is recalled. Eventually, in the last paragraph of this chapter, the idea for generalizing and improving the method is described.

\subsection{ECDLP}
\label{subsec:ecdlp}

Elliptic curve discrete logarithm problem is defined as finding such integer $y$, that for given elliptic curve $E$ over a prime field $\mathbb{F}_p$
\begin{equation}
	\label{eq:ecdlp1}
	[y]P = \underbrace{P+P+\dots+P}_{\text{y addends}} = Q,
\end{equation}

where $P$ and $Q$ are points on the curve $E$ and $y\in\left\lbrace 1,\dots,Ord(P)-1\right\rbrace$. Let $m$ be the bitlength of $Ord(P)$, then $y$ may be written with $m$ binary variables $u_i$ as
\begin{equation}
	\label{eq:y}
	y=2^{m-1}u_m+\dots+2u_2+u_1,
\end{equation}
what makes possible to transform Eq. \eqref{eq:ecdlp1} into 
\begin{equation}
	\label{eq:ecdlp_ext}
	\begin{array}{rl}
		Q=[y]P= &[2^{m-1}u_m+\dots+2u_2+u_1]P\\
		=&[2^{m-1}u_m]P+\ldots+[2u_2]P+[u_1]P\\
		=&[u_m]([2^{m-1}]P)+\ldots+[u_2]([2]P)+[u_1]P\\
		=&P_m+\ldots+P_2+P_1.
	\end{array}
\end{equation}

As an example, curve $E:y^2=x^3-3x+63$ over a prime field $\mathbb{F}_{1021}$ with order $ord(E)=964$ may be given. For points $P=(74,841)$ and $Q=[k]P=(1017,824)$, solving ECDLP means finding the proper $k$ value. For a finite field of a 10-bit length, dealing with this problem manually may be tedious, but any computer can accomplish it in fractions of a second and calculate the solution, which is $k=43$. The situation is quite different in modern cryptosystems, where the bit length of a finite field is expressed using hundreds of bits.

\subsection{QUBO}

Quadratic unconstrained binary optimization model is presented as
\begin{equation}
	\label{eq:qubo1}
	\min_{x\in \{0,1\}^N} x^T Q x,
\end{equation}
where $Q$ is an $N \times N$ upper-diagonal matrix of real weights, and $x$ is a vector of binary variables. It can be also defined as minimizing the function
\begin{equation}
	\label{eq:qubo2}
	f(x)=\sum_{i}{Q_{i,i} x_i}+\sum_{i<j}{Q_{i,j} x_i x_j}.
\end{equation}
As an example, a function $f(x_1,x_2,x_3)=3x_1^2-3x_3^2+5x_2x_3$ may be given. With the vector $x$ and the matrix $Q$, it may be written as
\begin{eqnarray*}
	\label{eq:qubo3}
	x=\left[\begin{array}{c}{x}_{1}\\ {x}_{2}\\ {x}_{3}\end{array}\right],\  Q=\left[\begin{array}{c c c}3 & 0 & 0 \\ 0 & 0 & 5\\ 0 & 0 & -3\end{array}\right].
\end{eqnarray*}
Minimal energy of the given function is $-3$ for $(x_1,x_2,x_3)=(0,0,1)$.

\subsection{Transformation}
\label{subsec:transform}

Since the starting and ending points for solving ECDLP have been described above, the last missing part is the road in between. It was minutely characterized in \cite{wronski2024transformation}. Below one can find a briefer explanation.

A single summand from the third line of Eq. \eqref{eq:ecdlp_ext} can be written as
\begin{equation}
	\label{ui_extend} 
	[u_i]\left([2^{i-1}]P \right) =
	\begin{cases}
		\mathcal{O} & \text{for }u_i=0,\\
		[2^{i-1}]P & \text{for }u_i=1.
	\end{cases}
\end{equation}
The two cases can be united into one expression as
\begin{equation}
	\label{eq:ui_short}
	[u_i]\left([2^{i-1}]P \right) = \mathcal{O}+u_i\left([2^{i-1}]P-\mathcal{O} \right) 
\end{equation}
and for the corresponding values of $u_i$ from Eq. \eqref{ui_extend}, the results are as assumed above. If the arithmetic of points on the chosen elliptic curve model is complete, the above formula may be divided into affine coordinates
\begin{equation}
	\label{eq:coordinates}		
	\begin{cases}
		P_{i,x}=\mathcal{O}_x+u_i\left([2^{i-1}]P_x-\mathcal{O}_x\right),\\
		P_{i,y}=\mathcal{O}_y+u_i\left([2^{i-1}]P_y-\mathcal{O}_y\right).
	\end{cases}
\end{equation}
Based on these equations, every precomputed multiplicity (only powers of 2) of point $P$ from Eq. \eqref{eq:ecdlp_ext} can be represented. In order to improve the intuition of the whole process, an illustration may be helpful.

\begin{figure}[h]
	\centering
	\includegraphics[width=0.6\linewidth]{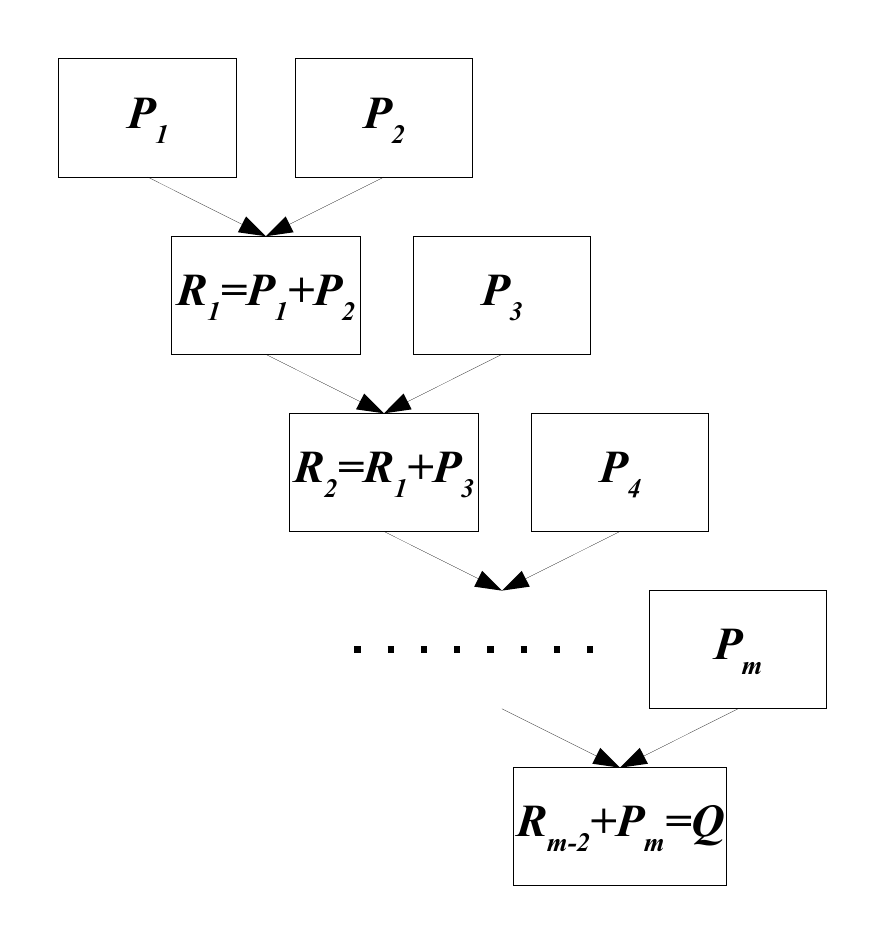}
	\caption{The decomposition of ECDLP \cite[Fig. 2]{wronski2024transformation}.}
	\label{fig:effappcur}
\end{figure}

The way of representing points $P_i$ is described above and the form of a given point $Q$ is with the given coordinates, just like in the example at the end of subsection \ref{subsec:ecdlp}. The only one unknown so far is the idea of representing the sum points $R_i$. Since ECDLP is given over some finite field, the coordinates of points $R_i$ must belong to this field. If so, each of them may be written with $n$ binary variables, where $n$ is the bit length of a field characteristic $p$.

Having knowledge about the representation forms of the points $P_i$, $R_i$ and $Q$, one may insert them into complete arithmetic equations on the selected elliptic curve model and perform the whole process. Well described example on the Edwards curve may be found in \cite{wronski2024transformation}. Moreover, in continuation of that research the authors have noticed, that changing the representation of a sought multiplicity of a point $P$ from binary to, for example, ternary may improve the process and lower the number of necessary resources. The results may be found in \cite{wronski2024base}.

However, there is still a problem: how to solve the ECDLP which is defined on curve models without complete arithmetic?

\subsection{Problems with the original method}

First of all, before the generalization is described, one needs to think about the motivation and ideas of how to go beyond the restrictions. The following conditions for solving ECDLP on the particular elliptic curve model have been mentioned by the authors of \cite{wronski2024transformation} for using the method from subsection \ref{subsec:transform}:
\begin{enumerate}
	\item arithmetic with a small number of multiplications,
	\item neutral element represented by affine coordinates,
	\item complete arithmetic.
\end{enumerate}

The first condition seems to affect the efficiency of the computations, so it may be omitted at the expense of increasing the number of necessary resources. 

The second prerequisite must be satisfied in the part of the existence of the numerical representation of the neutral element $\mathcal{O}$. Without that, points $P_i$ cannot be represented with the form from Eq. \eqref{eq:ui_short} and thus with Eqs. \eqref{eq:coordinates}. The type of coordinates influences mainly the efficiency, as, for example, computations in affine coordinates require less resources than in projective ones.

The last requirement forces using such arithmetic, that works correctly for all inputs. That means, it does not matter if the arguments are two different points, two same points or some point and the neutral point, for all these cases the formula should proceed and return a correct result. Due to the form of ECDLP decomposition in Eq. \eqref{eq:ecdlp_ext}, the arithmetic does not have to be unified (the same for adding and doubling points), because doubling points never happens, so adding one is enough. However, for the proper algorithm to work, the formula must take into account adding an arbitrary point with a neutral point $\mathcal{O}$.

Summing up, the biggest problem in generalizing the original method is the behavior of the algorithm when facing the neutral point. For elliptic curve models, in which the neutral point does not have a numerical form or the arithmetic does not cooperate with the neutral point (like in short Weierstrass model), the original method from \cite{wronski2024transformation} cannot work.

\subsection{Generalization - introduction}

To face that problem, the algorithm must be modified. The idea is to perform operations one level higher than in the original method. This means, that instead of converting points $P_i$ like in Eq. \eqref{eq:ui_short} and then inserting it into the addition formula, one may use the binary variables $u_i$ from Eq. \eqref{eq:y} and the third line of Eq. \eqref{eq:ecdlp_ext} and consider cases in addition $[u_i]P_i+[u_j]P_j$. 

The first one, when both $u_i$ and $u_j$ are positive. If so, the correct value is a point $P_i+P_j$.

The second one, when only $u_i$ is positive. Then, the proper result is $P_i$.

The third one, when only $u_j$ is positive. The correct result is $P_j$.

The last one, when neither $u_i$ nor $u_j$ is positive. Thinking logically, this is a case for a neutral point $\mathcal{O}$. However, it should not be used, because there may be an elliptic curve model, on which the corresponding arithmetic does not cooperate with the neutral points. For now, this case is unresolved but it will be below.

Based on the above considerations, the generalized addition may be written as
\begin{equation}
	[u_i]P_i+[u_j]P_j=
	\begin{cases}
		??? & \text{for } u_i=0\wedge u_j=0, \\
		P_i & \text{for } u_i=1\wedge u_j=0, \\
		P_j & \text{for } u_i=0\wedge u_j=1, \\
		P_i+P_j & \text{for } u_i=1\wedge u_j=1. \\
	\end{cases}
\end{equation}
As long as variables $u_i$ and $u_j$ are binary, the above cases may be transformed with boolean algebra into the expression
\begin{equation}
	\begin{array}{rl}
		[u_i]P_i+[u_j]P_j=&???+{[u_i(1-u_j)]P_i}+{[(1-u_i)u_j]P_j}
		+{[u_i\cdot u_j] (P_i+P_j)}.
	\end{array}
\end{equation}
Now the "???" signs have to be replaced with a numerical value. It can be done in at least two ways. Before the further description is presented, a few things should be noticed. The illustration in Fig. \ref{fig:effappcur} will be helpful.

Let's think about the first addition
\begin{equation*}
	P_1+P_2=[u_1]P+[u_2]\left([2]P \right).
\end{equation*}
If at least one out of the variables $u_i$ or $u_j$ is positive, the problematic case does not appear. What is more, that incident will not occur during subsequent additions, as all of the first arguments, being the results of all previous operations, will be different than the neutral element. Since binary variables in the first addition are the less significant bits of a multiplicity $y$, the analyzed "???" case appears when $y$ is divisible by 4.

\subsection{Generalization - first idea}
\label{subsec:1st}

The first idea of dealing with the problem is very simple. While solving ECDLP $[y]P=Q$, points $P$, $Q$ and the elliptic curve $E$ are known. If so, one may perform an addition $Q+P=[y+1]P$. Since the result of finding $y$ is not correct, it may indicate that $y$ is divisible by 4. If so, $y+1$ does not have this property so the algorithm should solve the problem. The small inconvenience is that the result must be reduced by 1. While using this solution, the case with $u_i$ and $u_j$ both negative can be omitted, so it is equal
\begin{equation}
	[u_i]P_i+[u_j]P_j=
	\begin{cases}
		P_i & \text{for } u_i=1\wedge u_j=0, \\
		P_j & \text{for } u_i=0\wedge u_j=1, \\
		P_i+P_j & \text{for } u_i=1\wedge u_j=1, \\
	\end{cases}
\end{equation}
and thus
\begin{equation}
	\label{eq:1st_add}
	\begin{array}{rl}
		[u_i]P_i+[u_j]P_j=&{[u_i(1-u_j)]P_i}+{[(1-u_i)u_j]P_j}
		+{[u_i\cdot u_j] (P_i+P_j)}.
	\end{array}
\end{equation}

\subsection{Generalization - second idea}

\begin{figure}[h]
	\centering
	\includegraphics[width=0.6\linewidth]{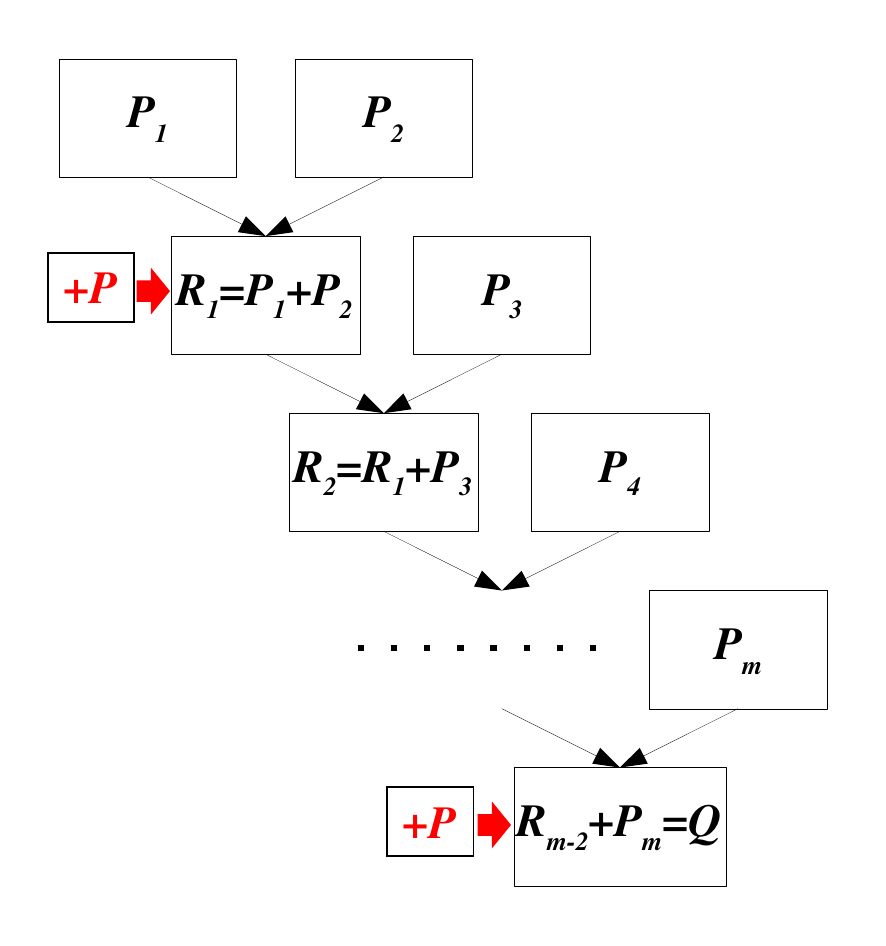}
	\caption{The modified decomposition of ECDLP.}
	\label{fig:effappcur22}
\end{figure}

The second idea is to conditionally add point $P$ during the first addition. Thanks to this, the result of a first operation will always be different than $\mathcal{O}$ and at the same time, potential problems in subsequent additions are eliminated. However, adding extra $P$ makes, that the solver has to deal with a problem $[y+1]P=[y]P$, which is not true if $P\neq\mathcal{O}$. That is why an extra $P$ has to be added to $Q$ as well. The modification was presented in the Fig. \ref{fig:effappcur22}.

Now the final version of arithmetic formulas may be presented. The first addition should be done with
\begin{equation}
	[u_i]P_i+[u_j]P_j=
	\begin{cases}
		P & \text{for } u_i=0\wedge u_j=0, \\
		P_i & \text{for } u_i=1\wedge u_j=0, \\
		P_j & \text{for } u_i=0\wedge u_j=1, \\
		P_i+P_j & \text{for } u_i=1\wedge u_j=1, \\
	\end{cases}
\end{equation}
what can be converted to
\begin{equation}
	\begin{array}{rl}
		[u_i]P_i+[u_j]P_j=&[(1-u_i)(1-u_j)]P+{[u_i(1-u_j)]P_i}\\
		&+{[(1-u_i)u_j]P_j}+{[u_i\cdot u_j] (P_i+P_j)}.
	\end{array}
\end{equation}
Due to the fact, that in subsequent additions the first argument is a sum of the previous elements and it is different than $\mathcal{O}$, the form of an operation $R_i+P_{i+2}$ is
\begin{equation}
	R_i+[u_{i+2}]P_{i+2}=
	\begin{cases}
		R_i & \text{for }u_{i+2}=0, \\
		R_i+P_{i+2} & \text{for }u_{i+2}=1, \\
	\end{cases}
\end{equation}
what is equal to
\begin{equation}
	R_i+[u_{i+2}]P_{i+2}={[(1-u_{i+2})]R_i}+{[u_{i+2}] (R_i+P_{i+2})}.
\end{equation}

\section{Practical example}

Because for fields of characteristic greater than 3 every elliptic curve can be transformed into a short Weierstrass curve, this model will be used in the example.

\subsection{Short Weierstrass curve}

Short Weierstrass curve $E_{SW}$ over prime field $\mathbb{F}_p$ is given by the equation
\begin{equation}
	y^2=x^3+ax+b,
\end{equation}
where $a,b$ are coordinates from $\mathbb{F}_p$.

The addition formulas on short Weierstrass curve $E_{SW}$ for points $P=(x_1,y_1)$, $Q=(x_2,y_2)$ and $R=P+Q=(x_3,y_3)$ are
\begin{equation}
	\label{eq:weier}
	\begin{array}{rl}
		x_3 &= \frac{(y_2-y_1)^2-(x_1+x_2)(x_2-x_1)^2}{(x_2-x_1)^2}=\frac{\text{nom}_x}{\text{denom}_x}, \\
		\\
		y_3 &= \frac{(2x_1+x_2)(y_2-y_1)(x_2-x_1)^2-(y_2-y_1)^3-y_1(x_2-x_1)^3}{(x_2-x_1)^3}
		= \frac{\text{nom}_y}{\text{denom}_y}.
	\end{array}
\end{equation}

\subsection{Explanation}

The example will be conducted with the first idea from Subsection \ref{subsec:1st}. To do this, Eq. \eqref{eq:1st_add} has to be used. Combining it with addition formulas Eqs. \eqref{eq:weier}, one obtains
\begin{equation}
	\label{eq:ri}
	\begin{array}{rl}
		R_i=&[u_i]P_i+[u_j]P_j=[u_i(1-u_j)]P_i+[(1-u_i)u_j]P_j+[u_i\cdot u_j] \frac{\text{nom}_{x/y}}{\text{denom}_{x/y}}.
	\end{array}
\end{equation}
Because in the QUBO model the equation must equal 0 and must not be represented with any fractions, the last transformation is required. After that, the formula is given by
\begin{equation}
	\label{eq:fi}
	\begin{array}{rl}
		F_i =& u_i\cdot u_j\cdot\text{nom}_{x/y}+\text{denom}_{x/y}\cdot \left( u_i\cdot(1-u_j)\cdot P_{i,x/y}\right.\\
		&\left.+u_j\cdot(1-u_i)\cdot P_{j,x/y}\right)-R_{i,x/y}\cdot \text{denom}_{x/y}.
	\end{array}
\end{equation}
Because the formula in Eq. \eqref{eq:fi} does not operate on points on an elliptic curve, instead uses polynomials and numerical coordinates, variables $u_i$ and $u_j$ are treated as integers from set $\left\lbrace 0,1 \right\rbrace$ and thus notation with square brackets is unnecessary and may be confusing. 

For every single addition, two functions $F_i$ will be obtained -- one for coordinate $x$ and one for coordinate $y$.

\subsection{Experiment}

Consider the following short Weierstrass curve $E_W/\mathbb{F}_3:y^2=x^3+3x+1$. The order of the group of points is equal to $7$ and the group is cyclic. The generator of this group is point $P= (2,1)$ and $Q= (0,2) = [y]P$. 

Since $ord(P)=7$ and $Q\neq\mathcal{O}$, $y\in\left\lbrace 1, \ldots, 6 \right\rbrace$. The sought multiplicity $y$ can be written as $y=4u_2+2u_1+u_0$. Based on \eqref{eq:ecdlp_ext}, one get
\begin{equation}
	\begin{array}{rl}
		[y]P&=[4u_2+2u_1+u_0]P=[4u_2]P+[2u_1]P+[u_0]P\\
		&=[u_2]\left([4]P\right)+[u_1]\left([2]P\right)+[u_0]P
		=
		[u_2]P_4+[u_1]P_2+[u_0]P_1. 
	\end{array}
\end{equation}
One can compute $P_4=\left(1,1 \right) $, $P_2=\left(0,1 \right) $ and $P_1=\left(2,1 \right) $. Because ECDLP is defined over $\mathbb{F}_3$ and $ord(P)=7$, only one point $R_i$ will be necessary and it has a form $R_1=\left(2u_4+u_5,\ 2u_6+u_7 \right) $.
From addition $[u_0]P_1+[u_1]P_2=R_1$ one obtains 
\begin{equation*}
	\begin{array}{rl}
		F_1 =&  2u_0u_1+2 u_0+u_4+2u_5, \\
		F_2 =& u_0+u_1+u_6+2u_7.
	\end{array}
\end{equation*}
From addition $R_1+[u_2]P4=Q$ one obtains
\begin{equation*}
	\begin{array}{rl}
		F_3 =& 2u_2u_4^3 + u_2u_5^3 +2 u_4^3 + u_5^3 + u_2u_6^2 + u_2u_6u_7 + u_2u_7^2\\ &+ u_4^2 + u_4u_5 + u_5^2 +2 u_2u_6 + u_2u_7 +2 u_4 + u_5, \\
		F_4 =&  u_2u_4^3 +2 u_2u_5^3 +2 u_4^3u_6 + u_5^3u_6 +2 u_2u_6^3 + u_4^3u_7\\ &+2 u_5^3u_7 + u_2u_7^3 + u_4^3 +2 u_5^3 +2 u_6 + u_7 + 1.
	\end{array}
\end{equation*}

Next, the following operations are performed:
\begin{enumerate}
	\item reducing $u^k$ to $u$ using a property of binary variables,
	\item transformation from the pseudo-boolean function over $F_3$ to the pseudo-boolean function over integers,
	\item linearization,
	\item summing squares of all equations.
	\item adding penalties.
\end{enumerate}
More details about the whole process can be found in \cite{wronski2022practical}.

\begin{figure}[h]
	\centering
	\includegraphics[width=0.4\linewidth]{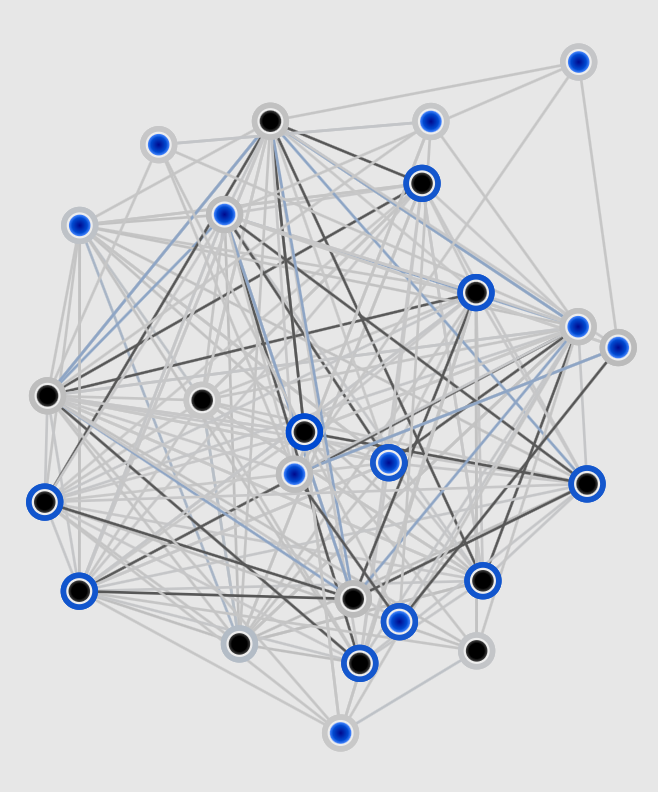}
	\caption{Connection between source variables.}
	\label{fig:dwave2}
\end{figure}

\begin{figure}[h]
	\centering
	\includegraphics[width=0.5\linewidth]{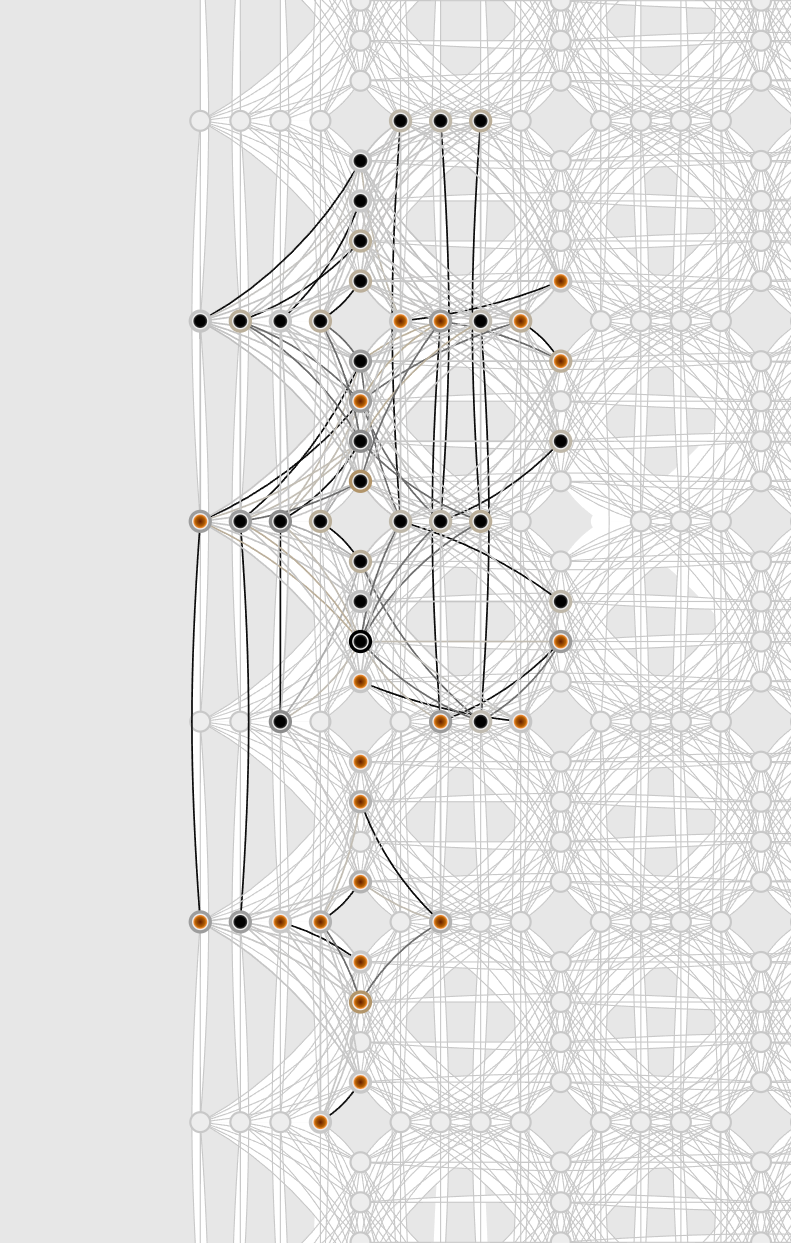}
	\caption{Embedding of a problem equivalent to the problem of finding elliptic curve discrete logarithm over $\mathbb{F}_3$ on shorts Weierstrass curve to the D-Wave Advantage.}
	\label{fig:dwave}
\end{figure}

The correct solution was found, which is $y=5$ $(u_2,u_1,u_0)=(1,0,1)$. The values of parameters used in solving this QUBO problem are shown in Tab. \ref{tab:params}. Connections between source variables are presented in Fig. \ref{fig:dwave2}. The embedding of a problem to D-Wave Advantage is presented in Fig. \ref{fig:dwave}.

\begin{table}[h]
	\label{tab:params}
	\centering
	\caption{D-Wave Advantage solver parameters used in solving QUBO problem equivalent to the problem of solving ECDLP over $\mathbb{F}_3$ on short Weierstrass curve in a subgroup of size 7}
	\begin{tabular}{|c|c|}
		\hline
		\textbf{Parameter}	&	\textbf{Value}  \\ \hline
		Solver	&  Advantage2\_prototype2.3 \\ \hline
		Qubits &	1248	\\ \hline
		Topology &	Zephyr	\\ \hline
		Number of read & 10000 	\\ \hline
		Annealing Time	&  200 \micro s \\ \hline
		Number of source variables & 25 \\ \hline
		Number of target variables & 51 \\ \hline
		Max chain length	& 3 \\ \hline
	\end{tabular}
\end{table}

\section{Conclusion and further work}

This paper presents a generalization of a method for solving the elliptic curve discrete logarithm problem with quantum annealing, described in \cite{wronski2024transformation}. The proper work of the method was confirmed with the practical implementation on the D-Wave computer with the D-Wave Leap cloud \cite{dwave}.

The ECDLP has been solved with a fully quantum device for a 2-bit finite field. Unfortunately, bigger problems have not been solved fully quantum. However, a few more problems have been solved with a hybrid solver. The biggest solved problem was ECDLP on an elliptic curve with order $7$ over a 4-bit field $\mathbb{F}_{11}$. Detailed information may be found in the below Tab. \ref{tab:wyniki}. About the colors of the cells in a table:
\begin{itemize}
	\item the green ones mean that the problem has been solved with a fully-quantum solver,
	\item the yellow ones mean that the problem has been solved with a hybrid solver,
	\item the white one that problem has not been solved.
\end{itemize} 

\begin{table}[h]
	\label{tab:wyniki}
	\caption{Number of binary variables necessary to solve ECDLP on elliptic curve with order 7}
	\centering
	\resizebox{0.7\columnwidth}{!}{%
		\begin{tabular}{|cccc|}
			\hline
			\multicolumn{1}{|c|}{Finite field bit length {[}b{]}} & \multicolumn{1}{c|}{Finite field}           & \multicolumn{1}{c|}{$1^{st}$ method}                    & $2^{nd}$ method                   \\ \hline
			\multicolumn{1}{|c|}{2}                        & \multicolumn{1}{c|}{$\mathbb{F}_3$}  & \multicolumn{1}{c|}{\cellcolor[HTML]{34FF34}26}  & \cellcolor[HTML]{34FF34}32 \\ \hline
			\multicolumn{1}{|c|}{3}                        & \multicolumn{1}{c|}{$\mathbb{F}_5$}  & \multicolumn{1}{c|}{\cellcolor[HTML]{F8FF00}74}  & \cellcolor[HTML]{F8FF00}98 \\ \hline
			\multicolumn{1}{|c|}{3}                        & \multicolumn{1}{c|}{$\mathbb{F}_7$}  & \multicolumn{1}{c|}{\cellcolor[HTML]{F8FF00}74}  & \cellcolor[HTML]{F8FF00}98 \\ \hline
			\multicolumn{1}{|c|}{4}                        & \multicolumn{1}{c|}{$\mathbb{F}_{11}$} & \multicolumn{1}{c|}{\cellcolor[HTML]{F8FF00}159} & 202                        \\ \hline
		\end{tabular}%
	}
\end{table}

The advantages of the generalized method:
\begin{itemize}
	\item does not require the numerical form of the neutral element,
	\item does not require using projective coordinates,
	\item allows to solve ECDLP on any model of elliptic curve,
	\item proper working is confirmed with quantum and hybrid computations, as well as with simulated annealing.
\end{itemize}

The disadvantages of the new method:
\begin{itemize}
	\item requires more resources than the original method.
\end{itemize}

As seen in Tab. \ref{tab:wyniki}, the first method requires fewer variables to solve ECDLP than the second one, however it demands supplementary actions for multiplicities divisible by 4. The alternative method does not need additional effort, but this comes at the cost of more necessary resources. Considering those arguments and the fact, that the extraordinary cases in a simpler method will happen with an expected probability about $25\%$, the author recommends using the first method. 

The description of the generalized method does not exhaust a topic. In further work, some improvements may be tried, for example:
\begin{itemize}
	\item changing the multiplicity representation like in \cite{wronski2024base},
	\item checking described method for other elliptic curve models,
	\item checking if for special cases using morphisms and arithmetic on other models will allow to use less resources like in \cite{wronski2016faster}.
\end{itemize}

\bibliography{dzierzkowski_general_ecdlp.bib}

%%%%%%%%%%%%%%%%%%%%%%%%%%%%%%%%%%%%%%%%%%%%%%%%%%%%%%%

\balance
\bibliographystyle{IEEEtran}
\end{document}